\documentclass[twocolumn]{aastex631}

\usepackage{wrapfig}
\usepackage{amsmath}
\usepackage{natbib}
\usepackage{graphicx}
\usepackage{appendix}
\usepackage{ulem}
\usepackage{soul}

\newcommand\msun{\hbox{\ensuremath{{\rm M}_{\odot}}}}

\newcommand{\mujy}{\mathrm{\ensuremath{\mu}Jy}}
\newcommand{\scinot}[1]{\ensuremath{\times 10^{#1}}}

\newcommand{\cliogal}{PEARLSDG}
\newcommand{\jwst}{{\it JWST}}

\begin{document}

\begin{abstract}
A wealth of observations have long suggested that the vast majority of isolated classical dwarf galaxies ($M_*=10^7$--$10^9$ M$_\odot$) are currently star-forming. However, recent observations of the large abundance of ``Ultra-Diffuse Galaxies" beyond the reach of previous large spectroscopic surveys suggest that our understanding of the dwarf galaxy population may be incomplete. Here we report the serendipitous discovery of an isolated quiescent dwarf galaxy in the nearby Universe, which was imaged as part of the \jwst{} PEARLS GTO program. Remarkably, individual red-giant branch stars are visible in this near-IR imaging, suggesting a distance of $30\pm 4$~Mpc, and a wealth of archival photometry point to an sSFR of $2\scinot{-11}$~yr$^{-1}$ and SFR of $4\scinot{-4}$~\msun{}~yr$^{-1}$. Spectra obtained with the Lowell Discovery Telescope find a recessional velocity consistent with the Hubble Flow and ${>}1500$~km/s separated from the nearest massive galaxy in SDSS, suggesting that this galaxy was either quenched from internal mechanisms or had a very high-velocity ($\gtrsim 1000$ km/s) interaction with a nearby massive galaxy in the past. This analysis highlights the possibility that many nearby quiescent dwarf galaxies are waiting to be discovered and that \jwst{} has the potential to resolve them.
\end{abstract}

%\title[]{PEARLS: A Deep JWST View of a Dwarf Spheroidal}
\title[]{PEARLS: A Potentially Isolated Quiescent Dwarf Galaxy with a TRGB Distance of $30$ Mpc}
\correspondingauthor{Timothy Carleton}
\email{tmcarlet@asu.edu}
\author[0000-0001-6650-2853]{Timothy Carleton} %%% tmcarlet@asu.edu
\affiliation{School of Earth and Space Exploration, Arizona State University,
Tempe, AZ 85287-1404, USA}

\author[0000-0001-5695-7002]{Timothy Ellsworth-Bowers} %%% 
\affiliation{Lowell Observatory, 1400 West Mars Hill Rd, Flagstaff AZ, 86001}

\author[0000-0001-8156-6281]{Rogier A. Windhorst}%%% Rogier.Windhorst@gmail.com
\affiliation{School of Earth and Space Exploration, Arizona State University,
Tempe, AZ 85287-1404, USA}

\author[0000-0003-3329-1337]{Seth H. Cohen} %%% seth.cohen@asu.edu
\affiliation{School of Earth and Space Exploration, Arizona State University,
Tempe, AZ 85287-1404, USA}

\author[0000-0003-1949-7638]{Christopher J. Conselice} %%% conselice@gmail.com 
\affiliation{Jodrell Bank Centre for Astrophysics, Alan Turing Building, 
University of Manchester, Oxford Road, Manchester M13 9PL, UK}

\author[0000-0001-9065-3926]{Jose M. Diego} %%% chemadiegor@gmail.com 
\affiliation{Instituto de F\'isica de Cantabria (CSIC-UC). Avenida. Los Castros
s/n. 39005 Santander, Spain}

\author[0000-0002-0350-4488]{Adi Zitrin} %%% 
\affiliation{Physics Department, Ben-Gurion University of the Negev, P.O. Box 653, Be’er-Sheva 84105, Israel}

\author[0000-0002-8449-4815]{Haylee N. Archer} %%% 
\affiliation{Lowell Observatory, 1400 West Mars Hill Rd, Flagstaff AZ, 86001}
\affiliation{School of Earth and Space Exploration, Arizona State University,
Tempe, AZ 85287-1404, USA}

\author[0000-0003-0230-6153]{Isabel McIntyre}
\affiliation{School of Earth and Space Exploration, Arizona State University,
Tempe, AZ 85287-1404, USA}

\author[0000-0001-9394-6732]{Patrick Kamieneski}
\affiliation{School of Earth and Space Exploration, Arizona State University,
Tempe, AZ 85287-1404, USA}

\author[0000-0003-1268-5230]{Rolf A. Jansen} %%% rolfjansen.work@gmail.com
\affiliation{School of Earth and Space Exploration, Arizona State University,
Tempe, AZ 85287-1404, USA}

\author[0000-0002-7265-7920]{Jake Summers} %%% jakesummers7200@gmail.com
\affiliation{School of Earth and Space Exploration, Arizona State University,
Tempe, AZ 85287-1404, USA}

\author[0000-0002-9816-1931]{Jordan C. J. D'Silva} %%% jordan.dsilva@research.uwa.edu.au
\affiliation{International Centre for Radio Astronomy Research (ICRAR) and the
International Space Centre (ISC), The University of Western Australia, M468,
35 Stirling Highway, Crawley, WA 6009, Australia}
\affiliation{ARC Centre of Excellence for All Sky Astrophysics in 3 Dimensions
(ASTRO 3D), Australia}

\author[0000-0002-6610-2048]{Anton M. Koekemoer} %%% koekemoer@stsci.edu
\affiliation{Space Telescope Science Institute,
3700 San Martin Drive, Baltimore, MD 21218, USA}

\author[0000-0001-7410-7669]{Dan Coe} %%% dcoe@stsci.edu
\affiliation{Space Telescope Science Institute, 3700 San Martin Drive, Baltimore, MD 21218, USA}
\affiliation{Association of Universities for Research in Astronomy (AURA) for the European Space Agency (ESA), STScI, Baltimore, MD 21218, USA}
\affiliation{Center for Astrophysical Sciences, Department of Physics and Astronomy, The Johns Hopkins University, 3400 N Charles St. Baltimore, MD 21218, USA}

\author[0000-0001-9491-7327]{Simon P. Driver} %%% Simon.Driver@icrar.org
\affiliation{International Centre for Radio Astronomy Research (ICRAR) and the
International Space Centre (ISC), The University of Western Australia, M468,
35 Stirling Highway, Crawley, WA 6009, Australia}

\author[0000-0003-1625-8009]{Brenda Frye} %%% brendafrye@gmail.com
\affiliation{Steward Observatory, University of Arizona, 933 N Cherry Ave,
Tucson, AZ, 85721-0009, USA}

\author[0000-0001-9440-8872]{Norman A. Grogin} %%% nagrogin@stsci.edu
\affiliation{Space Telescope Science Institute,
3700 San Martin Drive, Baltimore, MD 21218, USA}

\author[0000-0001-6434-7845]{Madeline A. Marshall} %%% madeline_marshall@outlook.com
\affiliation{National Research Council of Canada, Herzberg Astronomy \&
Astrophysics Research Centre, 5071 West Saanich Road, Victoria, BC V9E 2E7,
Canada}
\affiliation{ARC Centre of Excellence for All Sky Astrophysics in 3 Dimensions
(ASTRO 3D), Australia}

\author[0000-0001-6342-9662]{Mario Nonino} %%% nnn.mario@gmail.com
\affiliation{INAF-Osservatorio Astronomico di Trieste, Via Bazzoni 2, 34124
Trieste, Italy}

\author[0000-0003-3382-5941]{Nor Pirzkal} %%% npirzkal@stsci.edu
\affiliation{Space Telescope Science Institute,
3700 San Martin Drive, Baltimore, MD 21218, USA}

\author[0000-0003-0429-3579]{Aaron Robotham} %%% aaron.robotham@uwa.edu.au
\affiliation{International Centre for Radio Astronomy Research (ICRAR) and the
International Space Centre (ISC), The University of Western Australia, M468,
35 Stirling Highway, Crawley, WA 6009, Australia}

\author[0000-0003-0894-1588]{Russell E. Ryan, Jr.} %%% rryan.asu@stsci.edu
\affiliation{Space Telescope Science Institute,
3700 San Martin Drive, Baltimore, MD 21218, USA}

\author[0000-0002-6150-833X]{Rafael {Ortiz~III}} %%% rortizii@asu.edu
\affiliation{School of Earth and Space Exploration, Arizona State University,
Tempe, AZ 85287-1404, USA}

\author[0000-0001-9052-9837]{Scott Tompkins} %%% satompki@asu.edu
\affiliation{International Centre for Radio Astronomy Research (ICRAR) and the
International Space Centre (ISC), The University of Western Australia, M468,
35 Stirling Highway, Crawley, WA 6009, Australia}

\author[0000-0001-9262-9997]{Christopher N. A. Willmer} %%% cnawillmer@gmail.com
\affiliation{Steward Observatory, University of Arizona,
933 N Cherry Ave, Tucson, AZ, 85721-0009, USA}

\author[0000-0001-7592-7714]{Haojing Yan} %%% yanhaojing@gmail.com
\affiliation{Department of Physics and Astronomy, University of Missouri,
Columbia, MO 65211, USA}

\author[0000-0002-4884-6756]{Benne W. Holwerda} 
\affiliation{Department of Physics and Astronomy, University of Louisville, Louisville KY 40292, USA} 

\vspace{1em}

\section{Introduction}
\label{sec:intro}
Our understanding of the process of star formation and quenching in classical dwarf galaxies remains poorly understood, despite the large number of detailed observations of local systems \citep{Weisz2011,McConnachie2012,Spekkens2014,Putman2021,Mao2021,Carlsten2022}. This is partly due to the outsized influence of complex internal (e.g. star formation feedback; \citealt{Dekel2003,Hopkins2014,Agertz2016}) and external (e.g. ram-pressure stripping, galaxy harassment; \citealt{Gunn1972, Moore1996,Mayer2006,Fillingham2016, Boselli2008,Wang2021,Wang2022}) processes given their comparatively weak gravitational potential. These processes result in a large diversity in the star formation properties among the dwarf galaxy population \citep{Weisz2011,delosReyes2019}.

Despite all the variation in dwarf galaxy properties, one constant seems to hold: isolated dwarf galaxies always seem to be star-forming \citep{Haines2007,Geha2012,Kawinwanichakij2017,Davies2019,Prole2021}. Only a handful of objects are known to violate this rule \citep[e.g.][]{Karachentsev2015,Martinez-Delgado2016,Garling2020,Polzin2021,Casey2023}, and most of these objects are just beyond massive groups or clusters for which they may have experienced some recent interaction. However, observations of a large number of ``Ultra-Diffuse Galaxies" in clusters \citep{vanDokkum2015,Koda2015,Mihos2015,Munoz2015,Roman2017a,Lee2020}, groups \citep{vanderBurg2016}, and the field \citep{Leisman2017,Roman2017b,Prole2021}, have led some to speculate that the star-forming universality is hampered by selection effects and that many low surface brightness quiescent galaxies are waiting to be discovered \citep[e.g.~][]{Roman2019}. Results from the SMUDGES survey \citep{Zaritsky2019, Goto2023}, which finds a statistical signature of quiescent Ultra-Diffuse Galaxies well beyond the virial radii of massive hosts, give credence to this possibility.

Imaging with NIRCam \citep{Rigby2023} on \jwst{} has the potential to dramatically improve our understanding of nearby dwarf galaxy populations. Red-giant branch stars are approximately $2$ magnitudes brighter in the near-IR than optical wavelengths \citep{McQuinn2017,Weisz2023}, allowing for the possibility of measuring red giant branch distances beyond $30$~Mpc, and surface-brightness-fluctuation distances even further. This, in conjunction with the relative insensitivity of near-IR selected galaxies to age-based selection effects, means that a much more complete understanding of the environment of dwarf galaxies, and the influence of that environment on the star formation of those galaxies, will soon be possible.

As a precursor to this potential wealth of discovery, we report the serendipitous discovery of an isolated, quiescent, classical dwarf galaxy at RA=12h12m18s, Dec=+27d35m24s, known as \cliogal{} throughout, in imaging of the CLG1212 cluster as part of the Prime Extragalactic Areas for Reionization and Lensing Science (PEARLS) program \citep{Windhorst2023}. While this galaxy has been photometrically identified in other surveys (DECaLS and SDSS), \jwst{} imaging is able to resolve individual red-giant-branch stars constraining its distance to $30\pm4$~Mpc. Follow-up optical spectroscopy suggests that it is isolated from nearby massive galaxies and spectral energy distribution fitting confirms that it is quiescent. Section~\ref{sec:observations} describes \jwst{} and Lowell Discovery Telescope observations identifying the galaxy and measuring its recessional velocity. Section~\ref{sec:measurments} describes the measurement of its basic properties, including its recessional velocity, point-source photometry of its stars, and aperture photometry of the whole object. Section~\ref{sec:results} describes the inferred galaxy properties, including its distance measured with the TRGB method (Sec.~\ref{sec:dist}), its stellar population parameters and star formation rate based on spectral-energy-distribution fitting (Sec.~\ref{sec:sed}), and large-scale environment (Sec.~\ref{sec:env}). Finally, Section~\ref{sec:discussion} summarizes our results and presents some preliminary interpretations. We utilize Vega magnitudes when discussing point-source stellar photometry and Jy when discussing aperture photometry. When applicable, we utilize a cosmology with H$_0$=73~km~s$^{-1}$~Mpc$^{-1}$ \citep{Riess2022}, $\Omega_m=0.3$, and $\Omega_\Lambda=0.7$.  %Throughout, we refer to magnitudes using the AB system \citep{Oke1983}, utilizing a Vega-to-AB conversion from \cite{Willmer2018} when applicable. 

\section{Observations}
\label{sec:observations}
Before its serendipitous observation as part of the PEARLS program, \cliogal{} had been photometrically identified in SDSS, DECaLS \citep{Dey2019}, {\it WISE} \citep{Wright2010}, and {\it GALEX} surveys. It was also included in {\it Spitzer} IRAC $3.6~\mu$m, $4.5~\mu$m, $5.7~\mu$m, $8~\mu$m, and MIPS $24~\mu$m and $70~\mu$m imaging of CLG1212\footnote{It is just outside the footprint of $HST$ WFC3 and ACS imaging of the cluster taken as part of GO: 15959; PI: Zitrin.} (programs 20225, 13024; PI: Rines, Yan). For example, SDSS characterized it as an object with a $r$-band magnitude of $18.84$, a half-light radius of $3\farcs{}6$, and an average surface brightness of $23.6$~AB mag~arcsec$^{-2}$.  \jwst{} F200W observations find similar structural parameters, with a best-fit S{\'e}rsic $n$ of 0.8 and $r_e$ of $3\farcs{}7$ (see Sec.~\ref{sec:galfit}). This ancillary imaging allows us to characterize the stellar populations of \cliogal{} in detail.
 
\subsection{\jwst{} Observations}
\label{sec:jwstobs}
\begin{figure}
    \centering
    \includegraphics[width=1\linewidth]{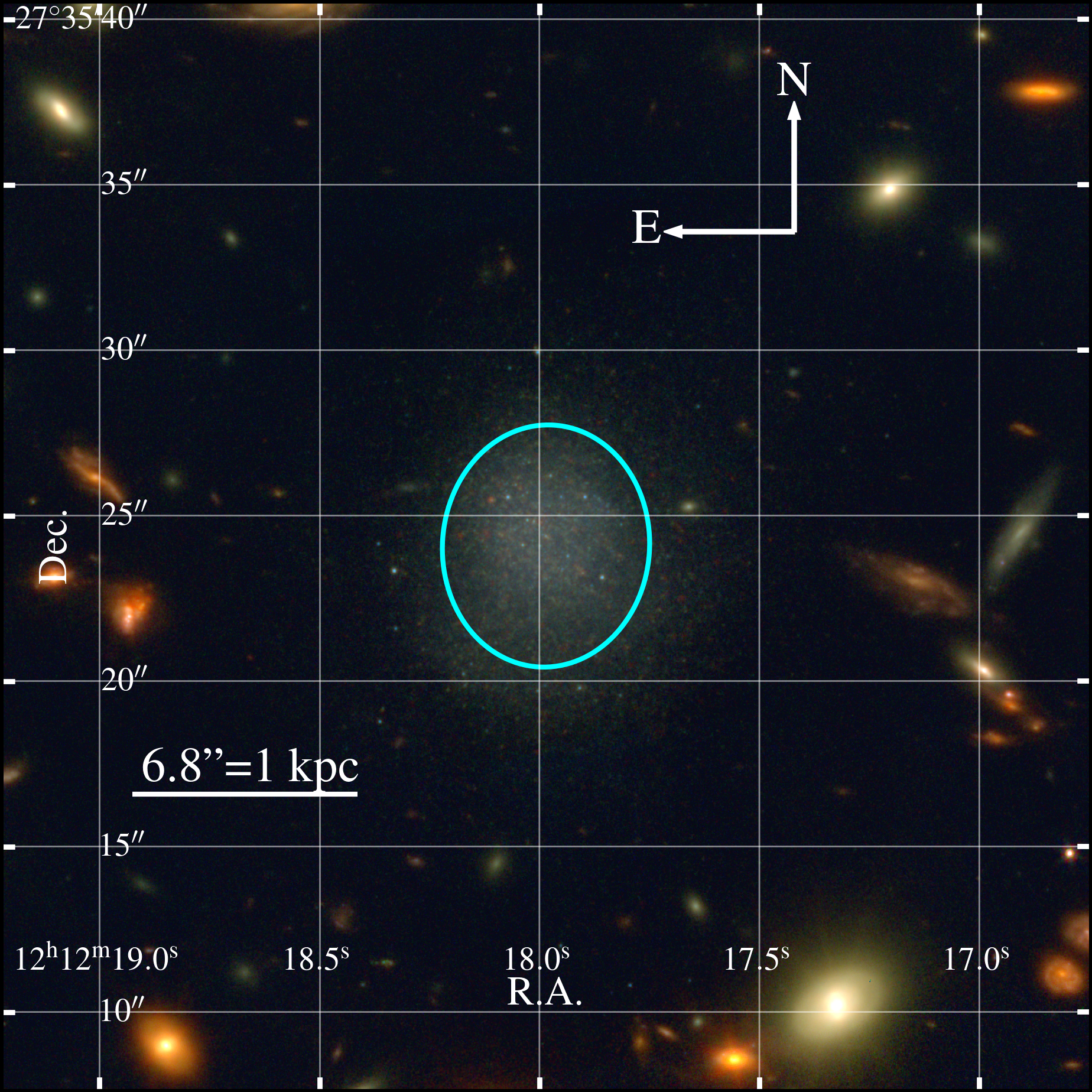}
    \includegraphics[width=1\linewidth]{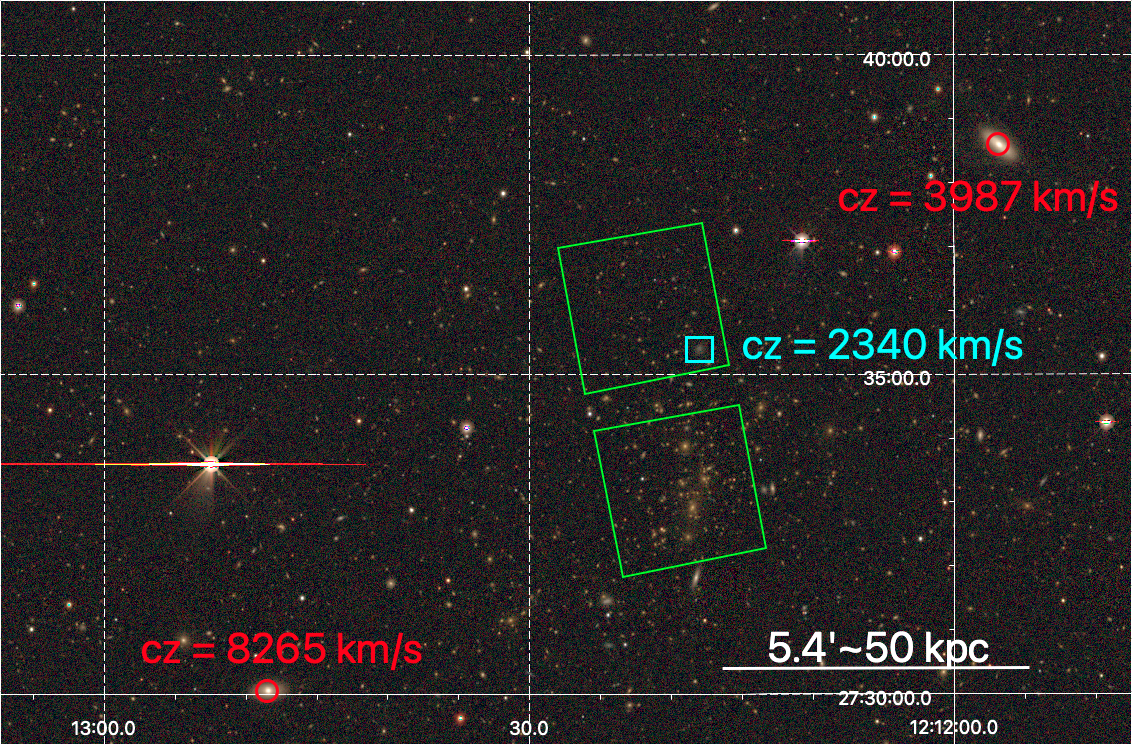}
    \caption{ {\bf Top:} The \jwst{} of the \cliogal{} galaxy (blue=F090W+F150W, green=F200W+0.5$\times$F277W, red=0.5$\times$F277W+F356W+F444W). {\bf Bottom:} DECALS $grz$ image of the sky immediately surrounding \cliogal{}. Both images are aligned such that North is up and East is left. \cliogal{} is identified with the cyan box, and the green squares show the area covered by NIRCam imaging. Also shown are two of the closest (in-projection) nearby massive galaxies (identified in red circles).}
    \label{fig:galimg}
\end{figure}

The Prime Extragalactic Areas for Reionization and Lensing Science PEARLS program (GTO 1176; PI Windhorst; \citealt{Windhorst2023}) targeted the CLG-J1212+2733 cluster \citep{Zitrin2020} on 2023 January 13-14.
This field was observed with F090W, F150W, and F200W short wavelength filters and F277W, F356W, and F444W long wavelength filters. The median exposure times were $2491$~s, $1890$~s, and $1890$~s for F090W, F150W, and F200W, and $1890$~s, $1890$~s, and $2491$~s for F277W, F356W, and F444W. In the imaging, \cliogal{} appears in the non-cluster module, approximately $2\farcm3$ from the cluster. Figure~\ref{fig:galimg}a shows an RGB image of \cliogal{} using all \jwst{} filters, and Figure~\ref{fig:galimg}b shows a DECaLS image of it and its surroundings.

The default PEARLS reductions described in \cite{Windhorst2023} apply ProFound-based sky-subtractions \citep{Robotham2017} and ``wisp" removal \citep{Robotham2023}, which was designed
to efficiently identify faint galaxies with small angular sizes.
%The default PEARLS reductions described in \cite{Windhorst2023} apply aggressive sky subtraction, with the goal of efficiently identifying distant galaxies with small angular sizes.
%Because of the excellent data quality associated with this reduction, we use it for the identification of stars (Sec.~\ref{sec:starmeasurement}).
However, this affects low surface brightness features in \cliogal{}, so we use the standard STSCI reductions, which do not implement this sky subtraction. The \jwst{} data are hosted at \dataset[DOI: doi:10.17909/h26w-zh06]{https://doi.org/doi:10.17909/h26w-zh06}, and will become publicly available 2024 January 13.

\subsection{Lowell Discovery Telescope DeVeny Observations}
\label{sec:ldtobs}
Following the identification of this galaxy, it was observed with the DeVeny long-slit optical spectrograph on the Lowell Discovery Telescope. The observations were carried out on 2023 June 21, using a 1\farcs5-wide slit and the 500~l/mm grating centered at $\lambda=5000$\AA. Eleven exposures were taken, with a total of 1.3 hours spent on source. Much of the spectrum is affected by sinusoidal pattern noise that can affect the Deveny camera\footnote{http://www2.lowell.edu/users/tbowers/DevenyManualv171.pdf}. This sinusoidal noise was first subtracted by fitting the pattern noise across the slit\footnote{https://github.com/LowellObservatory/LDTObserverTools}. Following this correction, standard data reductions were completed using the {\sc pypeit} software \citep{Prochaska2020}, which in addition to flat-field, bias, and wavelength calibrations, corrects for flexure effects using sky lines. The initial wavelength calibration was done using an ArI-CdI-Hg lamp, and sky lines were used to maintain the wavelength calibration throughout the night. The 2D spectra were stacked, weighting by the S/N of \cliogal{}, and 1D a spectrum was extracted using the optimal extraction procedure of \cite{Horne1986}.
%The one-dimensional spectra were extracted using optimal extraction, and stacked, using median stacking weighted by the S/N ratio.
%The 2D spectra were coadded, and the object spectrum was extracted using optimal extraction.
%The 1D extracted spectrum is shown as Fig.~\ref{fig:spectrum}. 

\section{Measurements}
\label{sec:measurments}
\subsection{Point-Source Photometry}
\label{sec:starmeasurement}
%To measure the distance and stellar populations of \cliogal, we first subtract the unresolved galaxy light. In each filter, we first mask bright objects using Source Extractor \citep{}. Then, we use Galfit \citep{} to fit the light profile of the galaxy. We find that two Sersic profiles fit the galaxy distribution best. The fit parameters are summarized in Table~\ref{tab:galfit}. We then conduct PSF photometry on the residual image. {\bf need to re-align pointing.} We model the PSF with WebbPSF. We use DOLPHOT to do the photometry (while it should be the same as photutils, the output quality checks are necessary and not included in photutils).
We conduct point-source photometry on \cliogal{} using the {\sc DOLPHOT} package \citep{Dolphin2000,Dolphin2016}. Updates to the {\sc DOLPHOT} software were implemented in April 2023 as part of the \jwst{} Resolved Stellar Populations Early Release Science Program \citep{Weisz2023}.
{\sc DOLPHOT} uses PSFs created with {\sc WebbPSF} to iteratively-subtract point sources identified in the image. Stars are identified in the combined {\sc i2d} file and simultaneously fit to the F090W, F150W, and F200W {\sc cal} files.
Aperture corrections are measured on isolated stars and applied to the measured fluxes. The parameters recommended for
\jwst{} observations in crowded fields\footnote{http://americano.dolphinsim.com/dolphot/dolphotNIRCAM.pdf} (including img\_apsky=20 35, img\_RAper = 3, and FitSky$=2$) were adopted. The drizzled F200W image (where red-giant branch stars are the brightest) was taken as the detection image, and photometry was conducted on all 6 \jwst{} filters.

Similar to other works, \citep[e.g.~][]{Dalcanton2009,Radburn-Smith2011,Danieli2020}, we limit our selection to objects with the following {\sc DOLPHOT} parameters: type$<=2$, S/N$_{F200W}>4$, S/N$_{F150W}>3$, S/N$_{F090W}>3$, {\sc crowd}$_{F200W}<0.3$, {\sc crowd}$_{F150W}<0.3$, {\sc crowd}$_{F090W}<0.3$, $|{\rm \sc sharp}|_{F200W}<0.2$,  $|{\rm \sc sharp}|_{F150W}<0.2$, $|{\rm \sc sharp}|_{F090W}<0.2$. We also exclude objects more than $8\arcsec$ from the galaxy center to reduce contamination from background point sources like globular clusters. The criteria of {\sc crowd}$<0.15$ largely restricts the sample to objects $>1.5\arcsec$ from the center of the galaxy, so we do not apply any additional spatial cut. 

\begin{figure*}
    \centering
    \begin{tabular}{m{.45\linewidth}c}
         \raisebox{+.5\height}{\includegraphics[width=1\linewidth]{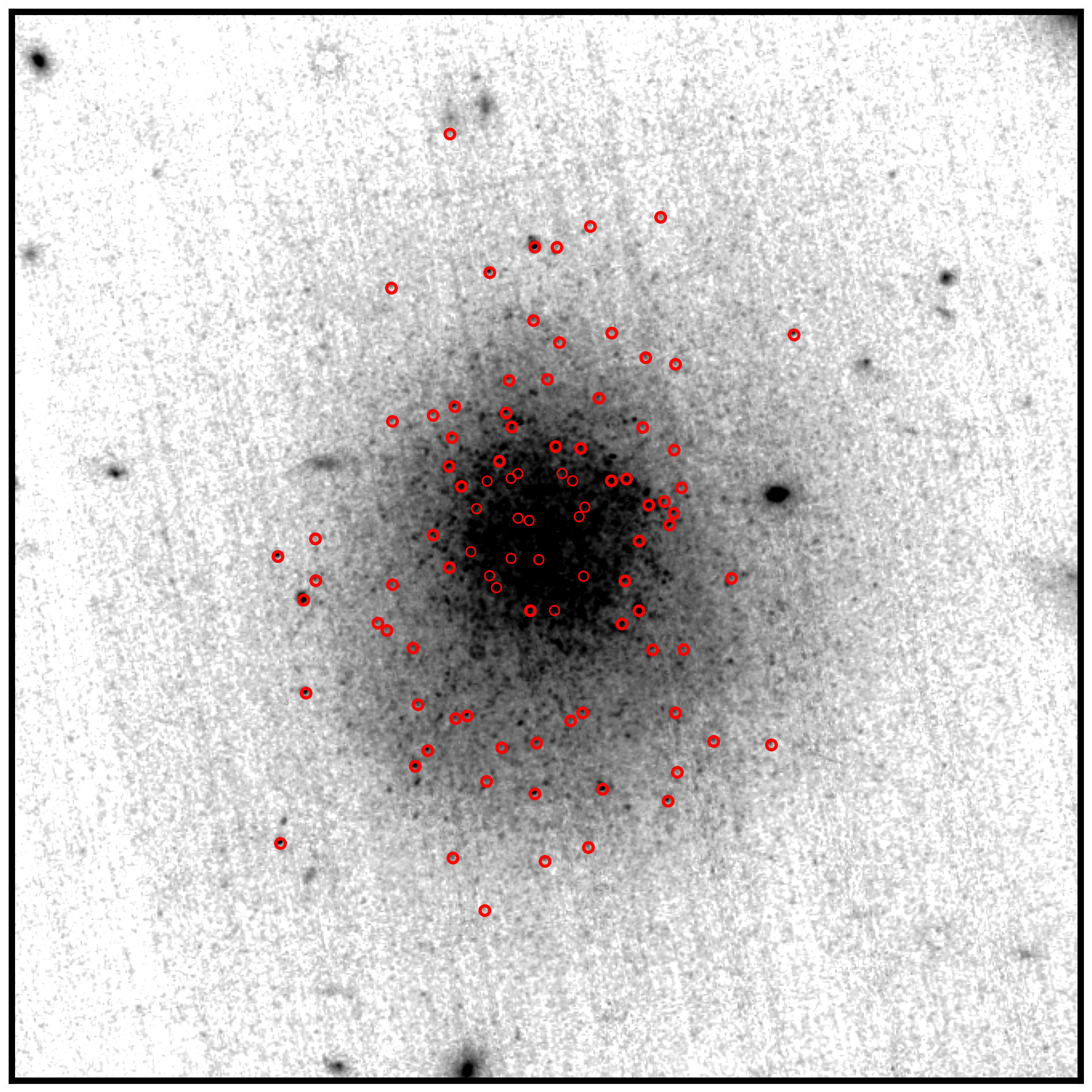}}
         &
         \includegraphics[width=.45\linewidth]{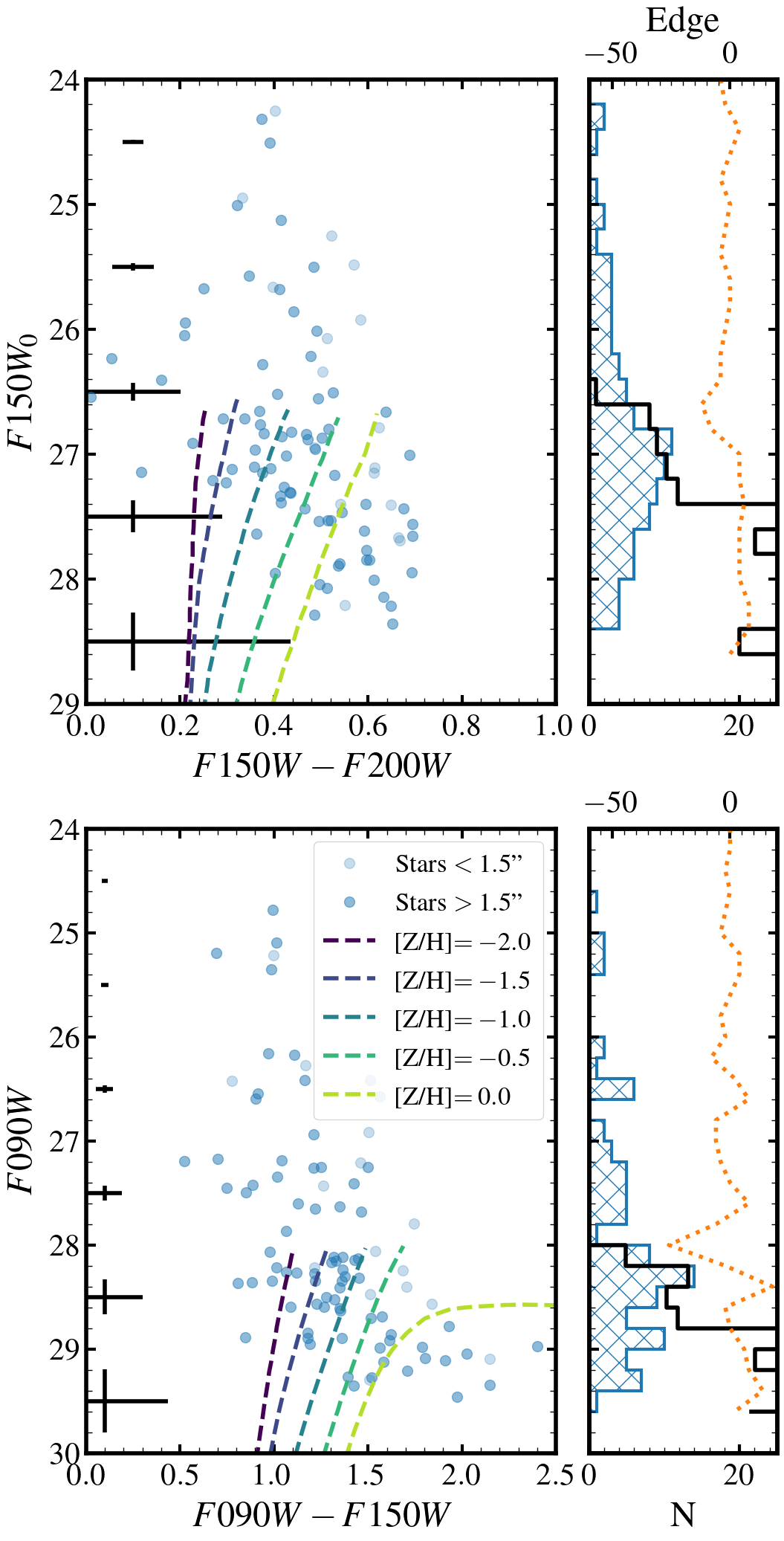}
    \end{tabular}
    \caption{{\bf Left:} F200W image of \cliogal{} with stars that pass our selection criteria circled in red (objects in the central $1.5\arcsec$ are thinner circles). While the imaging only pushes $\sim1$~mag below the TRGB, a number of red-giant branch stars are indeed visible. {\bf Right:} Color magnitude diagrams for $\rm F150W-F200W$ (top) and $\rm F090W-F150W$ (bottom) point-source photometry, with F150W$_0$ and F090W luminosity functions (far right). Objects in the central $1.5\arcsec$ are shown as lighter points. The $\rm F150W-F200W$ color-magnitude diagram has been rectified (see Sec.~\ref{sec:dist}). Average uncertainties as a function of magnitude are shown on the left. Also plotted are model $10$~Gyr old red-giant branch tracks of different metallicities (dashed lines) at a distance of $30.2$ Mpc. A number of stars with colors and magnitudes consistent with the brightest RGB stars are visible with $\rm F150W-F200W$ colors of 0.2--0.7. While our imaging is not deep enough to identify the RGB in F090W as clearly as F150W$_0$, bright RGB stars identified in F150W$_0$ and F200W are clearly detected in F090W\null. This allows us to fit the F090W luminosity function to determine the RGB tip and distance modulus. The model luminosity function (shown as the black line) matches the characteristic jump in the observed F090W and rectified F150W luminosity functions. Lastly, the Sobel Filter response is shown as the orange dotted line. The strongest peak matches the TRGB in F090W and F150W$_0$. These results illustrate the promise of \jwst{} to identify the TRGB in nearby galaxies.}
    \label{fig:cmag}
\end{figure*}

Additionally, several ($54$) stars have unexpectedly red colors, with F150W$-$F200W$>0.7$. The F090W$-$F150W colors of these objects are expected, but due to their unusual F150W$-$F200W colors, we exclude them from our sample.
This leaves us with $94$ stars.
Fig.~\ref{fig:cmag}a shows stars identified in the F200W image.
%As others have pointed out before \citep{Weisz2023,McQuinn2019}, \jwst{} has the potential to probe the red-giant branch to fainter magnitudes than before, given the cool temperatures of these stars. 

\subsection{Recessional Velocity}
\label{sec:redshift}
\begin{figure}
    \centering
    \includegraphics[width=1\linewidth]{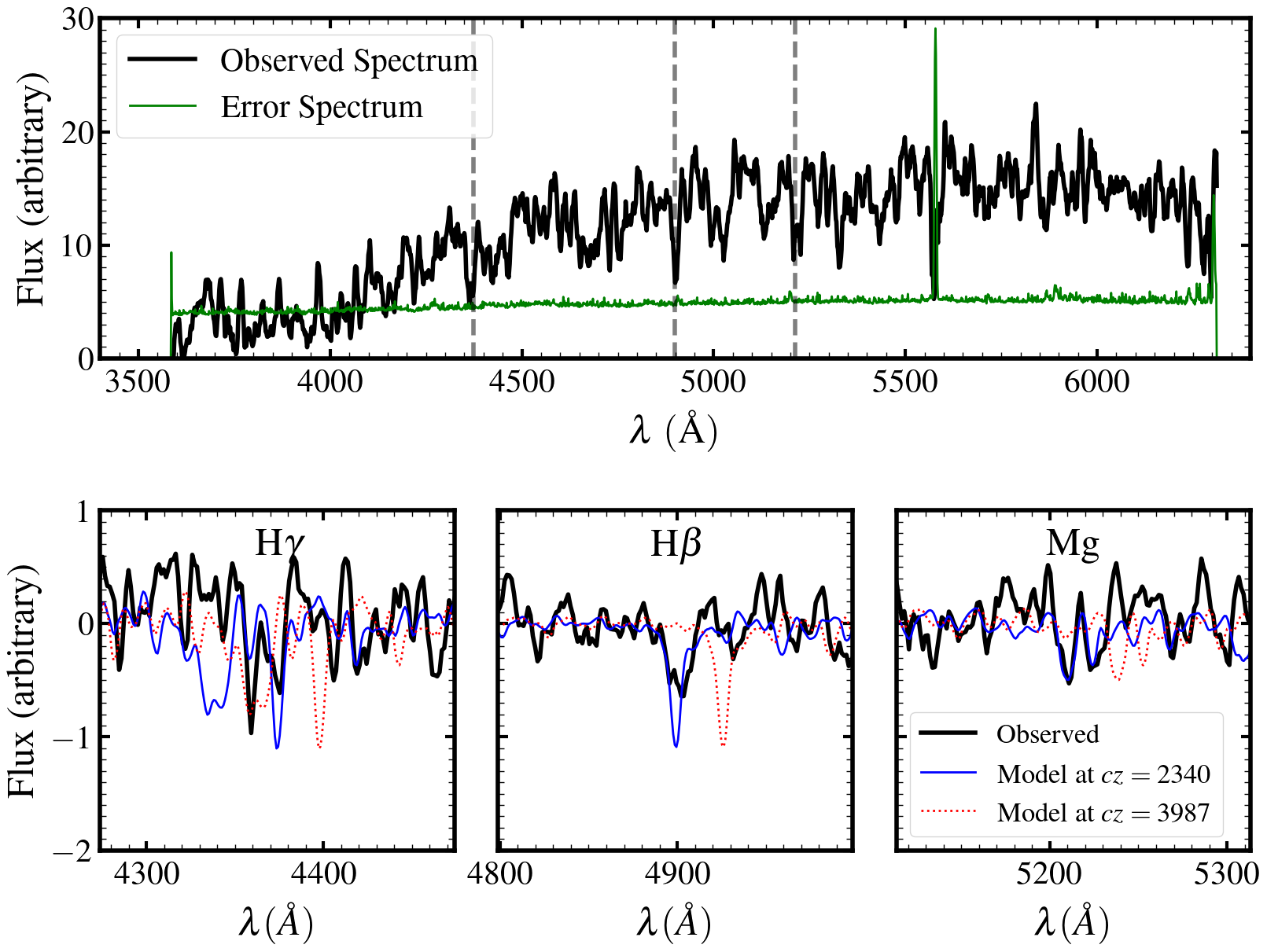}
    \caption{{\bf Top:} Observed spectrum (black) along with error spectrum (thin green line). {\bf Bottom:} Zoom in on the H$\gamma$, H$\beta$, and Mg features (highlighted by dashed lines in the top panel). The model spectrum at the measured recessional velocity of $cz=2340\pm180$~km/s is shown as the thin blue line. For comparison, the red dotted line shows the model spectrum redshifted to the nearest massive galaxy at $cz=3987$~km/s. \cliogal{} is $1650$~km/s separated from the nearest massive neighbor, so it is unlikely to be associated with it.}
    \label{fig:spectrum}
\end{figure}

As seen in Fig.~\ref{fig:spectrum}, the spectrum of \cliogal{} is relatively featureless and resembles a quiescent, low-mass galaxy.
%Unfortunately, much of the spectrum is affected by sinusoidal pattern noise that can affect the Deveny camera\footnote{This noise has been documented before (http://www2.lowell.edu/users/tbowers/DevenyManualv171.pdf), and it can sometimes be remedied by fitting this pattern noise across the slit. However, row-to-row phase variations are too large to correct for this noise in our observations.}.
While the spectrum is just above the sky background, at least three spectral features can be identified in the stacked, smoothed spectrum: H$\gamma$ at $\sim4370$\AA{}, H$\beta$ absorption at $\sim 4900$\AA{}, and Mg absorption at $5210$\AA{}.
%We use $25\AA$-wide windows on either side of those features to conduct a linear fit of the continuum across the line for each individual exposure. The residual spectra of all individual exposures are then stacked (weighted by the S/N of the individual observations).
%In the H$\beta$ and G-band features are visible, and marked in Fig.~\ref{fig:spectrum}.
To measure the recessional velocity of \cliogal{}, we cross-correlate a model spectrum (constructed with {\sc python FSPS} using a single stellar population of $10$~Gyr and metallicity of $-1.35$) with the observed stacked spectrum. We exclude wavelengths below $4250$\AA{} given the low S/N. The best-fit redshift is $z=0.0078$, corresponding to $2340\pm180$ km/s. The largest source of uncertainty in this measurement comes from the use of a wide slit to obtain high enough S/N, so we assign a recessional velocity error based on moving the center of the object halfway across the slit (taken at H$\beta$). This corresponds to a recessional velocity error of $180$ km/s.
%.75" = 2.21 pix = 2.93A = 0.0006 = 180 km/s
%In particular H$\beta$ is visible. To measure the central wavelength, we fit a gaussian to the line\footnote{We don't do a full spectral correlation measurement because the spectrum isn't flux-calibrated and there is some residual sinusoidal pattern noise.}.

\subsection{Aperture Photometry}
\label{sec:apphot}
To fully understand the stellar population properties of \cliogal{}, we conduct aperture-photometry on the existing UV-IR imaging, utilize archival imaging from $GALEX$, SDSS, DECals, \jwst{}, and $Spitzer$. First, we use Source Extractor \citep{Bertin1996} to identify and mask nearby galaxies. We expand the mask around all objects by 2 pixels to ensure we mask as much flux from these nearby galaxies as possible. Then, we convolve all images to the $4\farcs9$ resolution of $GALEX$.
The $GALEX$ and $Spizter$ PSFs were obtained online\footnote{http://www.galex.caltech.edu/researcher/techdoc-ch5.html and https://irsa.ipac.caltech.edu/data/SPITZER/docs/irac/ calibrationfiles/psfprf/}, the SDSS and DECALS PSFs were modeled as Gaussians with FWHM noted in the catalog data, and the \jwst{} PSFs were constructed from {\sc WebbPSF} v1.1.0.
With these convolved images, we conduct aperture photometry using python {\sc photutils} \citep{photutils}. Based on trial and error, we find that aperture-photometry with a $8\arcsec$ aperture minimizes the differences in measurements between different surveys and ensures we include nearly all the light from the galaxy in the convolved image. For \jwst{}, we use a $10-20\arcsec$ annulus (using the object-masked image) to estimate the background level; for SDSS and DECals, we utilize the existing background subtraction. For $GALEX$ we use the published background maps for background subtraction, and for $Spitzer$, we use the Source Extractor background maps. No $24~\mu$m emission is detected. Galactic extinction is corrected for using a \cite{Cardelli1989} extinction law assuming E(B-V)$=0.019$ \citep{Schlafly2011}. Note in Fig.~\ref{fig:galimg} that PEARLSDG is located in a sizeable NIRCam area that is fortuitously devoid of brighter objects, making aperture photometry and sky-subtraction possible in all these other images, which have much wider PSFs than \jwst{}.

\section{Results}
\label{sec:results}
\subsection{TRGB Distance}
\label{sec:dist}
%While the central region of \cliogal{} is crowded with stars, the outskirts are able to identify many individual red-giant-branch stars, as shown in The F150W-F200W vs. F150W color-magnitude diagram.
%This is verified by model RGBs from {\sc PARSEC} model stellar tracks \citep{Bressan2012,Marigo2013} set at a distance of $40$~Mpc (the distance to \cliogal{} assuming it is in the Hubble Flow). 
%{\bf redo with F090W RGB - RGB stars are detected at 3sigma in F090W, so we should use them}
The red-giant branch (RGB) ``tip," which represents the first He-flash of a large number of old red giant branch stars, has been used extensively to measure distances to nearby galaxies using optical measurements with $HST$ \citep[e.g.~][]{Salaris1997,Dalcanton2009, Dalcanton2012, Jang2017a,Jang2017b, Jang2020, McQuinn2017,Freedman2019,Danieli2020,Freedman2020}. The rectified F150W luminosity function shown Figure~\ref{fig:cmag} shows a distinctive discontinuity associated with this tip of the red-giant branch (TRGB).
Notably, this RGB tip is about $2$ magnitudes brighter in the near-IR compared with $I$-band \citep{McQuinn2019}, allowing it to be more easily identified in \jwst{} imaging.
However, while the structure of the RGB, and the absolute magnitude of the RGB tip have been shown to be insensitive to the parameters of the stellar population in $I$-band, the same is not necessarily true in the near-IR. In these wavelengths, the TRGB can vary by $0.75$ mag depending on the assumed metallicity. %Depending on the assumed stellar metallicity, the TRGB has a $\sim0.75$ magnitude spread.

Given that the TRGB is flat in F090W, we do not rectify the F090W$-$F150W color-magnitude diagram. On the other hand, the TRGB is expected to (and does in our data) have a slope in F150W$-$F200W. The number of stars in PEARLSDG is not enough to independently rectify this TRGB, so we fit a line to the TRGB of {\sc PARSEC} isochrones \citep{Bressan2012,Marigo2013} with metallicities of $-2.0$, $-1.5$, $-1.0$, $-0.5$, and $0$, all with a $10$~Gyr age to get the rectified F150W magnitude (which we refer to as F150W$_0$). We find a slope of $-2.66$ and we normalize to the TRGB color of the $-1.0$ metallicity track of $-0.392$.
To measure the TRGB distance, we take a forward-modeling approach following \cite{Danieli2020}. Given the proven calibration of the $I$-band TRGB and its insensitivity to metallicity, we utilize the F090W luminosity function to fit the TRGB and use the rectified F150W luminosity function as a check on this result.
%First, we select RGB and AGB stars in the F150W$-$F200W color-magnitude diagram using the boundary shown in Fig.~\ref{fig:cmag}. Next,
We generate an F090W luminosity function using the {\sc PARSEC} isochrones (with the metalicity set to Z/Z$_\odot=0.032$ and the age set to 10~Gyr) and a \cite{Kroupa2001} IMF. Then, we model the observed luminosity function as a combination of this luminosity function and contaminants:
\begin{equation}
dN(m)/dm = dN_{\rm track}(M+\mu)/dm + c_1 (m-27) + c_2,
\label{eqn:lfeqn}
\end{equation}
where the first term represents the modeled stellar population (primarily the RGB, but including AGB stars as well) shifted to the assumed distance modulus ($\mu$), and the second and third terms represent contamination (faint galaxies, pulsating AGB stars, foreground brown dwarfs, etc...). We optimize the likelihood of this model $100$ times, varying the individual star measurements by their photometric uncertainties, to estimate the range of allowed parameters. We optimize the model over $\mu$, $c_1$, and $c_2$, and find best-fit values of: $\mu=32.40\pm0.09$, $c_1=0.59\pm0.16$, and $c_2=1.36\pm0.2$.

%As a check on this result, we fit the F150W luminosity function with a similar procedure, and we find a $\mu=$, . 
%The {\sc PARSEC} models suggest that the TRGB in F090W should be at $x$ mag, right at the completeness limit of our F090W imaging. While we cannot fit the F090W TRGB to measure a distance less affected by systematics, we can use it to constrain the upper limit of the allowed distance modulus. Fitting the F090W luminosity function in a similar way as described above, we find a lower limit to the distance modulus of $32.5$.

This implies a distance of $30.2$~Mpc, consistent with its Hubble distance of {\bf $32\pm 2.5$}~Mpc. Although the statistical uncertainty of this measurement represents a $1.1$~kpc uncertainty, we adopt a $0.3$~mag, or $4$~kpc uncertainty to account for other uncertainties (e.g. in the TRGB calibration).
%To explore the range of distances allowed given the uncertain TRGB zero-point, we run this analysis using model stellar tracks ranging from Z/Z$_\odot=0.01$ to Z/Z$_\odot=1$. This gives a distance modulus range of $32.43$ to $32.47$, implying a distance range of $30.6-31.2$~Mpc. 
This measurement represents one of the most distant TRGB distance measurements to date \citep{Freedman2020} and highlights the potential that \jwst{} has to measure distances well beyond the local Universe.  

As a check on this modeling approach, we identify the tip of the red-giant branch by convolving the luminosity function with a Sobel filter ([-2,0,2]) for edge detection. (shown as the orange lines in Fig.~\ref{fig:cmag}). This finds a RGB tip at $28$~mag in F090W, within $0.15$ mag of the predicted TRGB at $30$~Mpc in F090W from \cite{McQuinn2019}. The tip of the rectified F150W$_0$ RGB is at $26.6$~mag, also consistent with the prediction from \cite{McQuinn2019}.

\subsection{Structural Parameters}
\label{sec:galfit}
To independently estimate the structural parameters of \cliogal{}, we fit the light profile with {\sc galfit} \citep{Peng2002}. We fit a single Sersic component and one sky component to the {\sc i2d} file downloaded from MAST. We include the derived parameters in Table~\ref{tab:sedresults}. Notably, \cliogal{} has a low Sersic index like many low surface brightness galaxies \citep[e.g.~][]{Yagi2016}.

%To model the red-giant branch and estimate the distance to \cliogal{}, we adopt two techniques. First we use a conservative approach, where stars along the line describing the red-giant-branch $\pm$ $0.x$ mag are selected, and a window function is used to identify the break in the star luminosity function associated with the tip-of-the-red-giant-branch. Second, we use all stars in the field. In the first case, we use an old, metal poor stellar population consistent with the red colors of \cliogal to model the red giant branch.

\subsection{SED Fitting}
\label{sec:sed}
\begin{figure}
    \centering
    \includegraphics[width=1\linewidth]{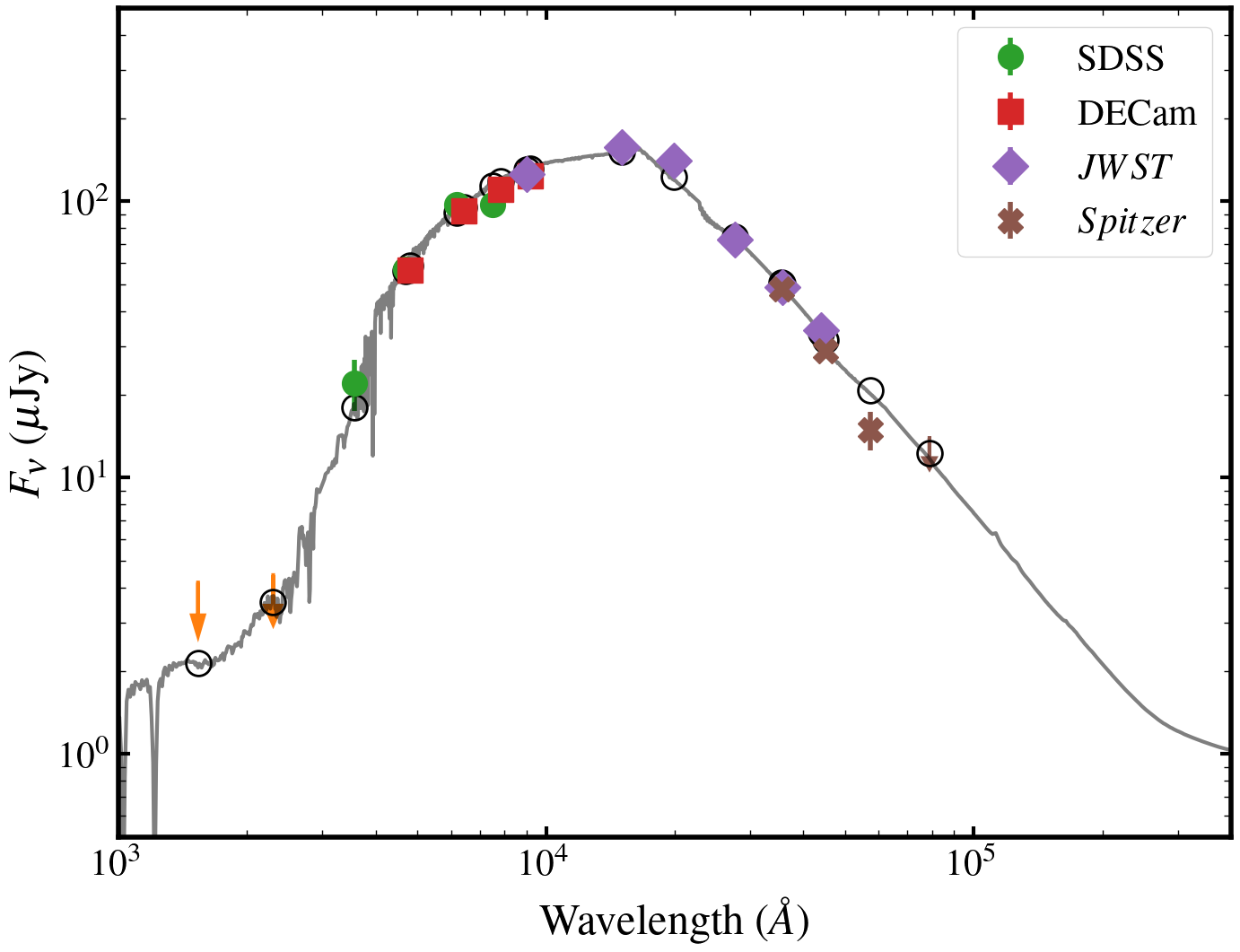}
    \caption{Broad-band spectral energy distribution of \cliogal{}, including all available UV, optical, and near-IR data, shown alongside the best-fit SED from {\sc Prospector}. The overlap of DECAM and SDSS, as well as \jwst{} and {\it Spitzer} bands support the accuracy of our photometry. The SED is fit well with a predominantly old, metal-poor stellar population, with a very small contribution from younger stars.}
    \label{fig:sed}
\end{figure}
As apparent given the lack of emission lines in the optical spectra, \cliogal{} does not have a high current star formation rate. To fully understand its stellar population properties we model its stellar population with {\sc Prospector} \citep{Johnson2021}, using the MILES stellar libraries \citep{Sanchez-Blazquez2006,Falcon-Barroso2011}, MIST isochrones \citep{Choi2016,Dotter2016}, \cite{Draine2007} dust templates, and a \cite{Kroupa2001} IMF. This model is fit to the aperture photometry described in Sec.~\ref{sec:apphot}. We adopt minimum uncertainties in the photometry of $1\%$ to account for systematic errors such as zero-point differences \citep{Rigby2023}. The measured and best-fit spectral energy distribution is shown in Fig.~\ref{fig:sed}. Despite the blue \jwst{} colors apparent in Fig.~\ref{fig:galimg}, the optical spectrum is very red ($FUV-r>3.4$), consistent with an old stellar population.

We model the star formation history as a 5-component star formation history with $t_{\rm age}/$yr bins at [[0,$10^{7.5}$],[$10^{7.5},10^{8.5}$],[$10^{8.5},10^{9.5}$],[$10^{9.5}$,$10^{10.11}$]]. In addition to modeling this star formation history, we fit for the dust content (simply modeled as a foreground screen with a power law with an index of $-0.7$ given its low star formation rate and low metallicity \citealt{Buzzo2022}).
%we test the impact of blue-horizontal-branch (bhb) stars on our fit by running the fit with the fraction of bhb stars of $0$, $0.25$, and $0.5$. We find that our results are not significantly impacted by $f_{bhb}$.
The results of our SED analysis are shown in Table~\ref{tab:sedresults}.
We find a best-fit metallicity of $-1.32$ and minimal dust extinction ($\tau_{V}$, the dust opacity at $5500$ has a best-fit value of $4.5\scinot{-4}$), consistent with its low mass and suggesting that it is not a tidal dwarf \citep[e.g.~][]{Duc2001}. Assuming the $30$~Mpc distance derived in Sec.~\ref{sec:dist}, we find a stellar mass of $1.7\scinot{7}$~\msun{} The $90\%$ upper limit on the fraction of the stellar population formed in the last $10^{8.5}$ years is $0.3\%$, and the best-fit sSFR (SFR) within that time is $2.4\scinot{-11}$~yr$^{-1}$ ($4\scinot{-4}$~\msun{}~yr$^{-1}$), with a $32-68$ percentile range of $1.7\scinot{-12}-1.4\scinot{-11}$~yr$^{-1}$ ($3\scinot{-5}-2.8\scinot{-4}$~\msun~yr$^{-1}$).

Statistical uncertainties in the inferred SED parameters are quite low given the precise photometry across a wide wavelength range. Modeling uncertainties, such as the assumed IMF, stellar libraries, and detector zero-points are likely the limiting uncertainties. How these translate to uncertainties in the inferred model parameters (the star formation history parameters in particular) is difficult to say. Attempting to fit the SED with other model assumptions results in similar results. For example, fitting with a delayed-$\tau$ star formation history finds a very low $\tau$ value and old age ($\tau\sim0.01$; $t_{\rm age}\sim10$~Gyr). This appears to be because the red near-IR colors (F090W through F200W) are only reproduced by a very old stellar population, although fitting the SED without including the F200W or F150W photometry still results in a low sSFR. Regardless, to be conservative we adopt a minimum $10\%$ systematic uncertainty in the inferred SED parameters, following \cite{Conroy2009}. 

The main uncertainty associated with the sSFR measurement is the amount of UV dust extinction. In our fiducial fitting, the low dust extinction is largely driven by the low $Spitzer$ $5$ and $8$ micron fluxes, although fitting with a PAH fraction of $10^{-4}$ still arrives at a best fit with very little dust extinction. Allowing preferential dust extinction around young ($10$~Myr old) similarly finds a low sSFR of $1.2\scinot{-11}$~yr$^{-1}$.
%Even allowing for a significant amount of dust extinction, the specific star formation rate within the last $30$ Myr is very low, consistent with a sSFR of $10^{-11}$~yr$^{-1}$. Some star formation within the last $300$~Myr is allowed, but still at a low rate of $\sim10^{-10}$~yr$^{-1}$.

Direct star formation rate estimates are not as constraining. The 3$\sigma$ $GALEX$ FUV luminosity is $4.8\scinot{24}$~erg~s$^{-1}$~Hz$^{-1}$. Following the calibration of \cite{McQuinn2015}, this results in an upper limit on the SFR of $9.8\scinot{-4}$~\msun{}~yr$^{-1}$. Adopting $40\%$ calibration uncertainty would put the limit at $1.4\scinot{-3}$~\msun{}~yr$^{-1}$. 

Regardless of the method used to estimate the SFR, we find that \cliogal{} has a remarkably low SFR. While constraints on the SFR-$M_*$ relation are sparse in this mass range, Local Volume dwarfs from \cite{Lee2011} have sSFRs of $7\scinot{-11}$~yr$^{-1}$, above all but our most conservative sSFR limit and $\sim0.5$~dex above our best estimate. Our best SFR estimate is $1.4$~dex below the SFR-M$_*$ relation of \citep{Salim2007} when extrapolated to the stellar mass of \cliogal{}. Lastly, we compare \cliogal{} to objects in the NASA Sloan Atlas\footnote{http://www.nsatlas.org/} (NSA, \citealt{Blanton2011}). Of objects with a stellar mass within $0.3$~dex within \cliogal{}, only $24\%$ have a lower sSFR and $21\%$ have a redder $FUV-r$ color.

\begin{table}[]
    \centering
    \begin{tabular}{l|r}
     Parameter &  Value \\
     \hline
     R.A. & 12h12m18s\\
     Dec. & +27d35m24s \\
     Distance & $30\pm4$ Mpc \\
     $r_h$ & $0.53\pm0.70$ kpc\\
     $b/a$ & $0.85\pm0.0007$ \\
     S{\'e}rsic $n$ & $0.79\pm0.0012$\\
     $f_{Galex~{\rm FUV}}$ & $<4.18~\mujy{}$ \\
     $f_{Galex~ {\rm NUV}}$ & $<4.44~\mujy{}$ \\
     $f_{{\rm SDSS}~ u}$ & $21\pm5.0~\mujy{}$\\
     $f_{{\rm SDSS}~ g}$ & $55.8\pm0.8~\mujy{}$ \\
     $f_{{\rm SDSS}~ r}$ & $95.6\pm0.9~\mujy{}$ \\
     $f_{{\rm SDSS}~ i}$ & $96\pm2.0~\mujy{}$ \\
     $f_{{\rm DECALS}~ g}$ & $55.2\pm0.8~\mujy{}$ \\
     $f_{{\rm DECALS}~ r}$ & $91.2\pm1.5~\mujy{}$ \\
     $f_{{\rm DECALS}~ i}$ & $110\pm1.9~\mujy{}$ \\
     $f_{{\rm DECALS~} z}$ & $123\pm3.0~\mujy{}$ \\
     $f_{JWST~ {\rm F090W}}$ & $125\pm1.2~\mujy{}$ \\
     $f_{JWST~ {\rm F150W}}$ & $156\pm1.6~\mujy{}$ \\
     $f_{JWST~ {\rm F200W}}$ & $140\pm1.4~\mujy{}$ \\
     $f_{JWST~ {\rm F277W}}$ & $72.6\pm0.7~\mujy{}$ \\
     $f_{JWST~ {\rm F356W}}$ & $48.9\pm0.5~\mujy{}$ \\
     $f_{JWST~ {\rm F444W}}$ & $34.1\pm0.3~\mujy{}$ \\
     $f_{Spitzer~ {\rm CH1}}$ & $48.1\pm0.4~\mujy{}$ \\
     $f_{Spitzer~ {\rm CH2}}$ & $28.9\pm0.7~\mujy{}$ \\
     $f_{Spitzer~ {\rm CH3}}$ & $15\pm2~\mujy{}$ \\
     $f_{Spitzer~ {\rm CH4}}$ & $<14~\mujy{}$ \\
     %{\bf write sfh in terms of percent of mass formed}
     Current total $M_*$ & $1.7\pm{0.2}\scinot{7}$~M$_\odot$\\
     $f_M*$ formed $\log~t_{\rm age}/{\rm yr}\in$ [0,7.5] & $1.6\pm^5_1\scinot{-5}$ \\
     $f_M*$ formed $\log~t_{\rm age}/{\rm yr}\in$ [7.5,8.5] & $5.2\pm^{10}_{5}\scinot{-4}$ \\
     $f_M*$ formed $\log~t_{\rm age}/{\rm yr}\in$ [8.5,9.7] & $2.9\pm^{30}_{3}\scinot{-5}$ \\
     $f_M*$ formed $\log~t_{\rm age}/{\rm yr}\in$ [9.7,10] & $9.2\pm^{120}_{9}\scinot{-5}$ \\
     $f_M*$ formed $\log~t_{\rm age}/{\rm yr}\in$ [10,10.11] & $0.997\pm{0.1}$ \\
     %$\log(Z/Z_\odot)$ & $-1.35\pm{0.02}$ \\
     $\log(Z/Z_\odot)$ & $-1.35\pm{0.1}$ \\
     $\tau_V$ & $1.4\pm^{0.94}_{0.79}\scinot{-3}$ \\
    \end{tabular}
    \caption{Information about \cliogal{}. The half-light radius ($r_h$), axis ratio ($b/a$) and S{\'e}rsic index (S{\'e}rsic $n$) are derived from Galfit modeling of the F200W image. Individual fluxes are calculated following the procedure of Sec~\ref{sec:apphot} with a minimum $1\%$ uncertainty. The fraction of stellar mass formed in various age bins ($f_M*$), metallicity ($Z/$Z$_\odot$), and dust opacity at $5500$~\AA{} ($\tau_V$) are derived from Prospector SED fitting. The uncertainties are probably too small given possible systematic uncertainties, but it is clear that this galaxy is predominantly composed of an old, metal-poor stellar population.}
    \label{tab:sedresults}
\end{table}

\subsection{Environment}
\label{sec:env}
\begin{figure}
    \centering
    \includegraphics[width=1\linewidth]{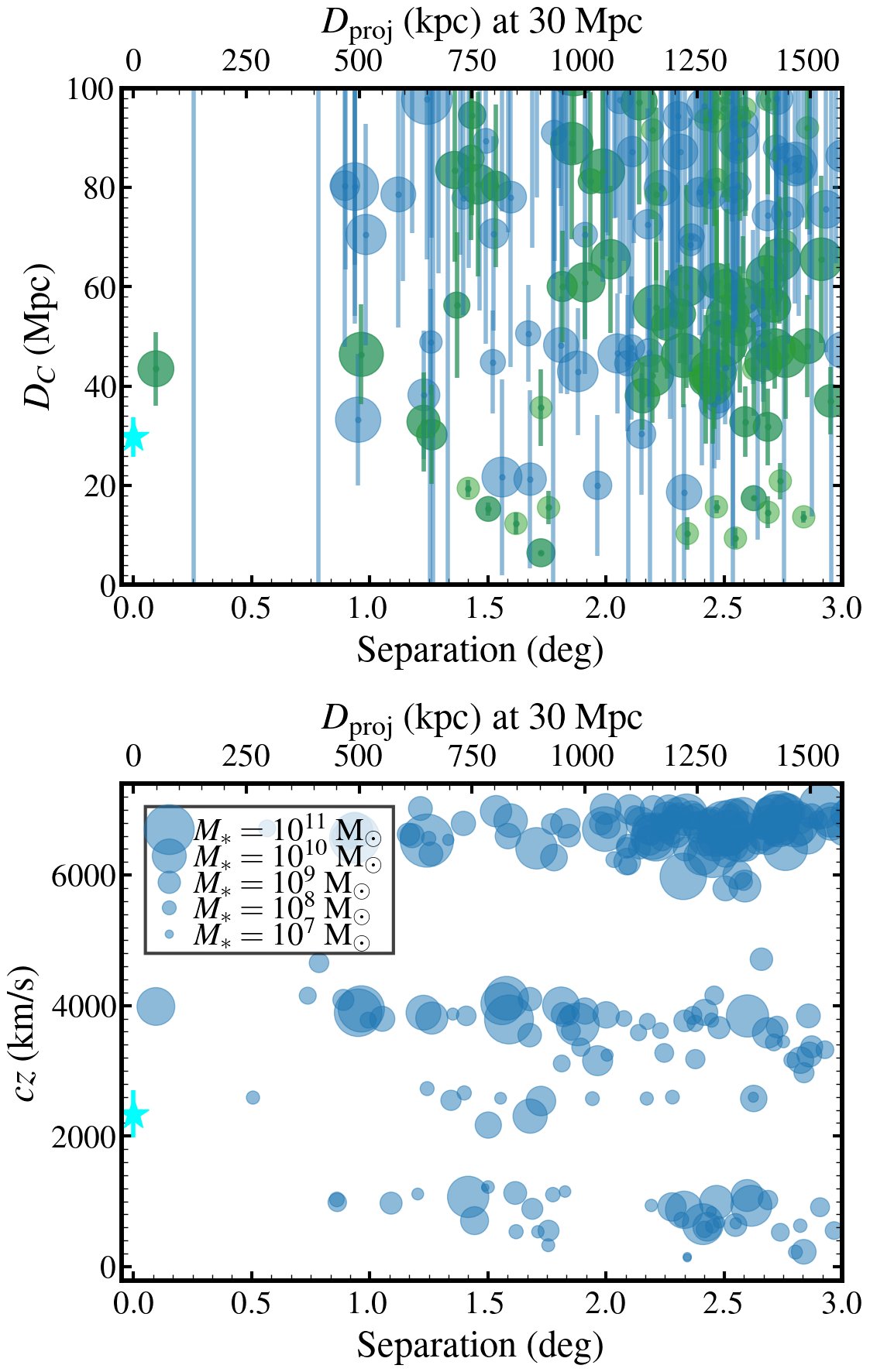}
    \caption{The large-scale environment of \cliogal{}. Both plots show the location of nearby galaxies with respect to \cliogal{} (shown as the cyan star), with point sizes proportional to the stellar mass of the system. {\bf Top:} The comoving distance to \cliogal{} from its TRGB, as a subset of objects in the NSA with either direct distance estimates from Cosmicflows-4 (green points) or our Fundamental Plane analysis (blue points). Objects in the Cosmicflows-4 catalog without a counterpart in the NSA are shown as light green points defaulting to a stellar mass of $10^9$~\msun{}. {\bf Bottom:} The recessional velocity of \cliogal{} and nearby systems in the NSA. \cliogal{} appears to be beyond the sphere of influence of any massive galaxy.}
    \label{fig:env}
\end{figure}

To probe the environment of this galaxy, we draw from the NASA Sloan Atlas \citep{Blanton2011}. This catalog is optimized to analyze nearby galaxies in SDSS, such as \cliogal{} and its neighbors. We supplement this data with distance estimates from the CosmoFlows-4 catalog \citep{Tully2023}, containing direct distance estimates for a large number of local galaxies. Figure~\ref{fig:env} shows \cliogal{} in the context of its surroundings, both in projected distance vs. luminosity distance space (using the Fundamental Plane \citealt{Cappellari2013} for galaxies besides \cliogal{} and those with direct distance estimates from Cosmicflows-4) and projected distance vs. recessional velocity space (bottom). While \cliogal{} is the same general RA, DEC coordinates of Virgo, Coma, and the Great Wall, it is actually in a very isolated region of space. 

The closest massive galaxy (SDSS J121156.80+273835.5, or J1227) is $1650$~km/s separated from \cliogal{}, and there are no massive ($>10^9$~\msun{}) galaxies within $1000$~km/s and $1$~Mpc, making it one of the most isolated quiescent dwarf galaxies observed. This is further demonstrated in the top panel of Fig.~\ref{fig:env}; the CosmicFlows-4 distance to J1227 is $43.5\pm 7$~Mpc, $1.9\sigma$ away from \cliogal{}. This is in agreement with the flow-model distance from Cosmicflows-4\footnote{https://edd.ifa.hawaii.edu/CF4calculator/} \citep{Kourkchi2020}, and would require a $+1797$~km/s peculiar velocity to be nearby in a region of space where the typical peculiar velocity is $-195$~km/s, further suggesting that \cliogal{} and J1227 are indeed not physically associated.

Regardless, we cannot completely rule out past interactions with other galaxies that may have affected its formation history. For example, it is possible it had a high-speed interaction with J1227 recently, and was quenched by that flyby interaction \citep{Benavides2021}. Alternatively, perhaps it interacted with nearby low-mass galaxies or a cosmic sheet and was quenched through that interaction \citep{Garling2020,Pasha2023}. However, the recessional velocity and luminosity distance of \cliogal{} are consistent with it being in the Hubble Flow, and there are no visible signatures of tidal interactions (see Fig.~\ref{fig:galimg}).

\section{Discussion}
\label{sec:discussion}
In this paper, we have reported the serendipitous discovery of \cliogal{}: a dwarf galaxy in PEARLS imaging of the CLG1212 field. This deep \jwst{} imaging allows us to resolve individual RGB stars in this object and characterize its distance as $30\pm4$~Mpc. This represents one of the furthest objects for which a TRGB distance has been determined and highlights the potential for \jwst{} to measure distances to galaxies in the nearby Universe. 

By combining PEARLS imaging with existing UV-IR imaging, we are able to constrain the stellar population properties of \cliogal{}. Consistent with its low level of UV emission and the lack of emission lines in its spectrum, we find a very low sSFR, suggesting that its star formation shut off over $1$~Gyr ago. Deeper follow-up spectroscopy is necessary to understand its formation history and abundance patterns in detail.

Most models for quenching dwarf galaxies have focused on environmental effects \citep{Bahe2015,Alberts2022} such as ram-pressure stripping \citep[e.g.~][]{Gunn1972,Fillingham2016,Bekki2009,Boselli2022}, strangulation \citep{Larson1980,Kawata2008}, or tidal stripping \citep{Moore1996}. However, recent observations of large numbers of Ultra-Diffuse Galaxies have prompted the development of internal quenching mechanisms, such as strong feedback \citep{Mori2002}. More unusual environmental effects such as flyby quenching, in which a quenched galaxy is ejected from the host after a high speed interaction, have also been proposed \citep{Benavides2021}. More detailed analysis of the star formation history of \cliogal{} and the dynamics of \cliogal{} with respect to its surroundings are needed to further understand its formation history, but this discovery suggests the possibility that many isolated quiescent galaxies are waiting to be identified and that \jwst{} has the tools to do so.

\vspace{1em}
\noindent{\bf Acknowledgements:}
%%% \begin{acknowledgements} %%% has page formatting problems
%
This work is based on observations made with the NASA/ESA/CSA James Webb Space
Telescope. The data were obtained from the Mikulski Archive for Space Telescopes
(MAST) at the Space Telescope Science Institute, which is operated by the
Association of Universities for Research in Astronomy, Inc., under NASA contract
NAS 5-03127 for {\it JWST}. These observations are associated with {\it JWST} programs 1176 and 2738.

TMC is grateful for support from the Beus Center for Cosmic Foundations. RAW, SHC, and RAJ acknowledge support from NASA {\it JWST} Interdisciplinary
Scientist grants NAG5-12460, NNX14AN10G and 80NSSC18K0200 from GSFC. JMD acknowledges the support of project PGC2018-101814-B-100 (MCIU/AEI/MINECO/FEDER, UE) Ministerio de Ciencia, Investigaci\'on y Universidades.  This project was funded by the Agencia Estatal de Investigaci\'on, Unidad de Excelencia Mar\'ia de Maeztu, ref. MDM-2017-0765. CC is supported by the National Natural Science Foundation of China, No. 11803044, 11933003, 12173045. This work is sponsored (in part) by the Chinese Academy of Sciences (CAS), through a grant to the CAS South America Center for Astronomy (CASSACA). We acknowledge the science research grants from the China Manned Space Project with NO. CMS-CSST-2021-A05. RAB gratefully acknowledges support from the European Space Agency (ESA) Research Fellowship. CJC acknowledges support from the European Research Council (ERC) Advanced Investigator
Grant EPOCHS (788113). CNAW acknowledges funding from the {\it JWST}/NIRCam contract NASS-0215 to the University of Arizona. MAM acknowledges the support of a National Research Council of Canada Plaskett Fellowship, and the Australian Research Council Centre of Excellence for All Sky Astrophysics in 3 Dimensions (ASTRO 3D), through project number CE17010001.

We also acknowledge the indigenous peoples of Arizona, including the Akimel
O'odham (Pima) and Pee Posh (Maricopa) Indian Communities, whose care and
keeping of the land has enabled us to be at ASU's Tempe campus in the Salt
River Valley, where much of our work was conducted. Lowell Observatory sits at the base of mountains sacred to tribes throughout the region. We honor their past, present, and future generations, who have lived here for millennia and will forever call this place home. 

\software{ Astropy: \url{http://www.astropy.org} \citep{Robitaille2021,
Astropy2018}; Photutils:
\url{https://photutils.readthedocs.io/en/stable/} \citep{Bradley2020};
Profound: \url{https://github.com/asgr/ProFound} \citep{Robotham2017, 
Robotham2018}; ProFit: \url{https://github.com/ICRAR/ProFit}
\citep{Robotham2018}; SourceExtractor:
\url{https://sextractor.readthedocs.io/en/latest/} \citep{Bertin1996}; Python FSPS: \url{https://dfm.io/python-fsps/current/} \citep{Conroy2009,Conroy2010,Foreman-Mackey2014}; Prospector: \url{https://prospect.readthedocs.io/en/latest/} \citep{Johnson2021}; WebbPSF: \url{https://webbpsf.readthedocs.io/en/latest/}.}
\facilities{James Webb Space Telescope; Mikulski Archive
\url{https://archive.stsci.edu}; Lowell Discovery Telescope.}

\bibliography{main}

\end{document}